\documentclass[fleqn,usenatbib]{mnras}

\usepackage{amssymb}

\usepackage{newtxtext,newtxmath}

\usepackage[T1]{fontenc}

\DeclareRobustCommand{\VAN}[3]{#2}
\let\VANthebibliography\thebibliography
\def\thebibliography{\DeclareRobustCommand{\VAN}[3]{##3}\VANthebibliography}

\usepackage{graphicx}	
\usepackage{comment}

\usepackage[usenames,dvipsnames]{xcolor}

\newcommand{\Teff}{\mbox{$T_\mathrm{eff}$}}
\newcommand{\logg}{\mbox{$\log g$}}

\newcommand{\Porb}{\mbox{$P_\mathrm{orb}$}}
\newcommand{\obj}{WD\,0141--675}

\newcommand{\Ion}[2]{#1{\,\sc#2}}
\newcommand{\kms}{\mbox{$\mathrm{km\,s^{-1}}$}}

\title[ESPRESSO observations of WD\,0141--675]{ESPRESSO observations of the debris-accreting white dwarf WD\,0141--675}

\author[Ram\'irez et al.]{
Sergio H. Ram\'irez$^1$\thanks{E-mail: sergio.ramirez.astro@gmail.com}, 
Boris T. G\"ansicke$^{1,2}$,
Detlev Koester$^3$,
Marina Lafarga$^{1,2}$, Nicola P. Gentile-Fusillo$^{4}$\\
$^{1}$Department of Physics, University of Warwick, Coventry, CV4 7AL, UK\\
$^{2}$Centre for Exoplanets and Habitability, University of Warwick, Coventry, CV4 7AL, UK\\
$^{3}$Institut für Theoretische Physik und Astrophysik, University of Kiel, 24098 Kiel, Germany\\
$^{4}$Department of Physics, Universit\`a degli Studi di Trieste, Via A. Valerio 2, 34127, Trieste, Italy}

\date{Accepted XXX. Received YYY; in original form ZZZ}

\pubyear{2021}

\begin{document}
\label{firstpage}
\pagerange{\pageref{firstpage}--\pageref{lastpage}}
\maketitle

\begin{abstract}
\obj\ was reported as the first astrometrically detected white dwarf planet host candidate as part of the third data release from \textit{Gaia}, just to be later retracted via a news item on the \textit{Gaia} web site$^1$. We present time-resolved, high-resolution optical ESPRESSO spectroscopy of \obj. A radial velocity analysis of the \Ion{Ca}~K absorption line reveals a tentative periodic signal of $15.6\pm0.9$\,d and an amplitude modulation of $0.19\pm0.11$\,km\,s$^{-1}$.  Phase-folding the ESPRESSO spectroscopy on this signal exhibits weak variability in the morphology of \Ion{Ca}~K close to the core of the line. A violet-to-red ratio analysis of the \Ion{Ca}~K line shows a periodic signal of $16.7\pm1.0$\,d. The periods from both methods agree, within their uncertainties, with half the period of the astrometric planet candidate, however, both measurements are of low statistical significance. Nonetheless, our results imply possible solutions to the mass function within the planetary regime. When combined with existing infrared photometry, which rules out a brown dwarf companion, yield a lower limit on the orbital inclination of $\sim7^\circ$.
Our study demonstrates that ESPRESSO observations are well capable of detecting short-period (days to weeks) giant planets orbiting white dwarfs. 
\end{abstract}

\begin{keywords}
techniques: radial velocities -- planets and satellites: composition -- planets and satellites: detection -- stars: individual: WD\,0141--675 -- white dwarfs.
\end{keywords}



\section{Introduction}
Most planet hosts will eventually evolve into white dwarfs, and large parts of their planetary systems are expected to survive that metamorphosis \citep{duncan+lissauer98-1, debesetal02-1, villaver+livio07-1, veras+gaensicke15-1}. Rich but indirect observational evidence for such evolved planetary systems around white dwarfs has been found in the form of circumstellar dust \citep{zuckerman+becklin87-1, farihietal09-1} and gas \citep{gaensickeetal06-3} discs, the detection of photometric transits from debris \citep{vanderburgetal15-1, vanderboschetal20-1}, and, most abundantly, metal-enrichment of the white dwarf atmospheres caused by the accretion of planetary debris \citep{zuckermanetal03-1, koesteretal14-1}. However, details on the architectures of the planetary systems around white dwarfs remain uncertain.

Early searches for transiting planets orbiting white dwarfs were unsuccessful \citep{burleighetal02-1, burleighetal08-1, mullallyetal08-1, kilicetal09-1}. A planet-mass common proper motion companion to a white dwarf, WD\,0806--661b, was found with an orbital separation of $\simeq2500$\,au \citep{luhmanetal11-1}, raising questions about its formation mechanism (\citealt{rodriguezetal11-1}, see also \citealt{verasetal09-1}). Recent spectroscopic \citep{gaensickeetal19-1} and photometric \citep{vanderburgetal20-1} observations of white dwarf planets  showed not only that they can survive the giant phase of their host stars, but can also migrate onto close-in ($\simeq0.02$\,au) orbits. A giant planet on wider ($\sim$\,au) orbits has been detected using micro-lensing \citep{blackmanetal21-1};  \textit{JWST} imaging identified planet candidates with projected separations of $\simeq12$ and $\simeq35$\,au at two nearby white dwarfs \citep{mullallyetal24-1}; and \textit{JWST} spectroscopy found evidence for a planet candidate at the nearby WD\,0310$-$688 via the detection of an infrared excess \citep{limbachetal24-1}. Whereas these candidates require confirmation, evidence collected so far suggests that giant planets orbiting white dwarfs may be rather common. 

\textit{Astrometry} provides another method to identify planets orbiting nearby white dwarfs, and the potential of the \textit{Gaia} mission in that respect has been thoroughly explored \citep{silvottietal11-1, sandersonetal22-1}. Indeed, as part of its third data release (Gaia DR3; \citealt{gaiaetal23-1, holletal23-1}), the first astrometric white dwarf planet candidate was announced: WD\,0141--675.

\section{WD\,0141--675}
WD\,0141--675 is a nearby ($d=9.71$\,pc, \citealt{gaiaetal23-2}) white dwarf with a hydrogen-dominated atmosphere \citep{hintzen+jensen79-1}, with weak Ca~H\&K absorption lines \citep{debes+kilic10-1}, indicating ongoing accretion of planetary material. Interest in WD\,0141--675 was recently sparked when an astrometric planet candidate with an orbital period of $P=33.65\pm0.05$\,d and a mass of $M_\mathrm{c}=9.26{+2.64 \atop -1.15}\,\mathrm{M_{Jup}}$\ was announced \citep{gaiaetal23-1, holletal23-1}. However in 2023 May, the planet candidate was quietly retracted via a news item\footnote{\url{https://www.cosmos.esa.int/web/gaia/dr3-known-issues}}, and classified as a false positive in the \textit{Gaia} astrometric solutions.

\citet{rogersetal24-1} presented a prescription for investigating the nature of astrometric candidate planets using \obj~ as a case study. These authors stress the importance of follow-up spectroscopic observations to confirm a planetary companion. With this in mind, we obtained high-resolution spectroscopy of \obj\ using ESPRESSO (Echelle SPectrograph for Rocky Exoplanets and Stable Spectroscopic Observations) on the Very Large Telescope (VLT), to probe for radial velocity variations indicative of a planetary companion. ESPRESSO is an optimal spectrograph for such objective, as was made evident by \citet{pasquinietal-2023}, who reported highly-accurate velocity shifts of eight Hyades white dwarfs using this instrument, showcasing its capabilities. \par
In this paper we report the results of our time-resolved spectra spanning the full putative orbital period of  $\Porb=33.65\pm0.05$\,d. In particular, we focus on the \ion{Ca}\,H\&K absorption lines, which are the only features in the spectrum of \obj\ apt for high-resolution orbital and morphological variation analyses, due to their sharp and narrow structure.

\section{Observations}
We acquired high-resolution spectroscopy of WD\,0141--675 with the ESPRESSO spectrograph at the VLT of the European Southern Observatory (ESO). We used the single-telescope configuration, slow read-out and $4\times2$ binning ($\mathrm{spatial}\times\mathrm{spectral}$). We obtained a total of 20 spectra, each with an exposure time of 3144\,s, throughout the period 2023 October 10 to 2023 December 20; further details of the observations are provided in Table\,\ref{t-obslog}. The observations were designed to sample the putative orbital period of  $\Porb=33.65\pm0.05$\,d, obtaining one spectrum on average once every two days. However, due to varying sky conditions, several nights were skipped, and the observations ended up being spread out over two months.

The ESPRESSO data were reduced with the Data Reduction Software\footnote{\url{www.eso.org/sci/software/pipelines/espresso/espresso-pipe-recipes.html}} (DRS) version 3.0.0. The DRS performs standard echelle spectrum reduction, including bias and dark subtraction, optimal order extraction, bad pixel correction, flat-fielding, deblazing, wavelength calibration, and order merging \citep[for details see][]{pepe2021espresso}. Here we used the deblazed, order-merged spectra. We removed telluric lines using {\sc molecfit} \citep{kauschetal15-1,smetteetal15-1}. The signal-to-noise ratio (S/N) of the individual spectra, measured around the  \ion{Ca}~K absorption line, ranges from 14 to 25 (Table\,\ref{t-obslog}).

\section{Atmospheric analysis}
We used the atmosphere code of \citet{koester10-1} to model the Ca\,K line in the ESPRESSO spectra of WD\,0141$-$675. Because ESPRESSO is designed primarily for high-precision velocity measurements, the flux calibration is not sufficiently accurate over wavelength scales comparable to the width of the Balmer lines to attempt a spectroscopic fit of the atmospheric parameters, i.e. the white dwarf effective temperature, \Teff, and its surface gravity, \logg. We therefore adopted the values from \citet{rogersetal24-1}, who derived spectroscopic and photometric parameters of ($\Teff=6421\pm90$\,K, $\log g=8.10\pm0.04$) and ($\Teff=6321\pm50$\,K, $\log g=7.97\pm0.03$), respectively. Fitting the Ca abundances in the white dwarf atmosphere, we find $\log(\mathrm{Ca/H})=-10.75\pm0.05$ for the spectroscopic parameters, and $\log(\mathrm{Ca/H})=-10.85\pm0.05$ for the photometric ones, in agreement with the results of \citet{rogersetal24-1}. Our atmosphere models included all elements up to Zn, with the exception of B, F, and Ar, scaled relative to the Ca abundance adopting a bulk Earth composition. No additional metal absorption lines are predicted by the model within the wavelength range covered by the ESPRESSO data, and none were detected.

\begin{figure}
\includegraphics[width=1\columnwidth,]{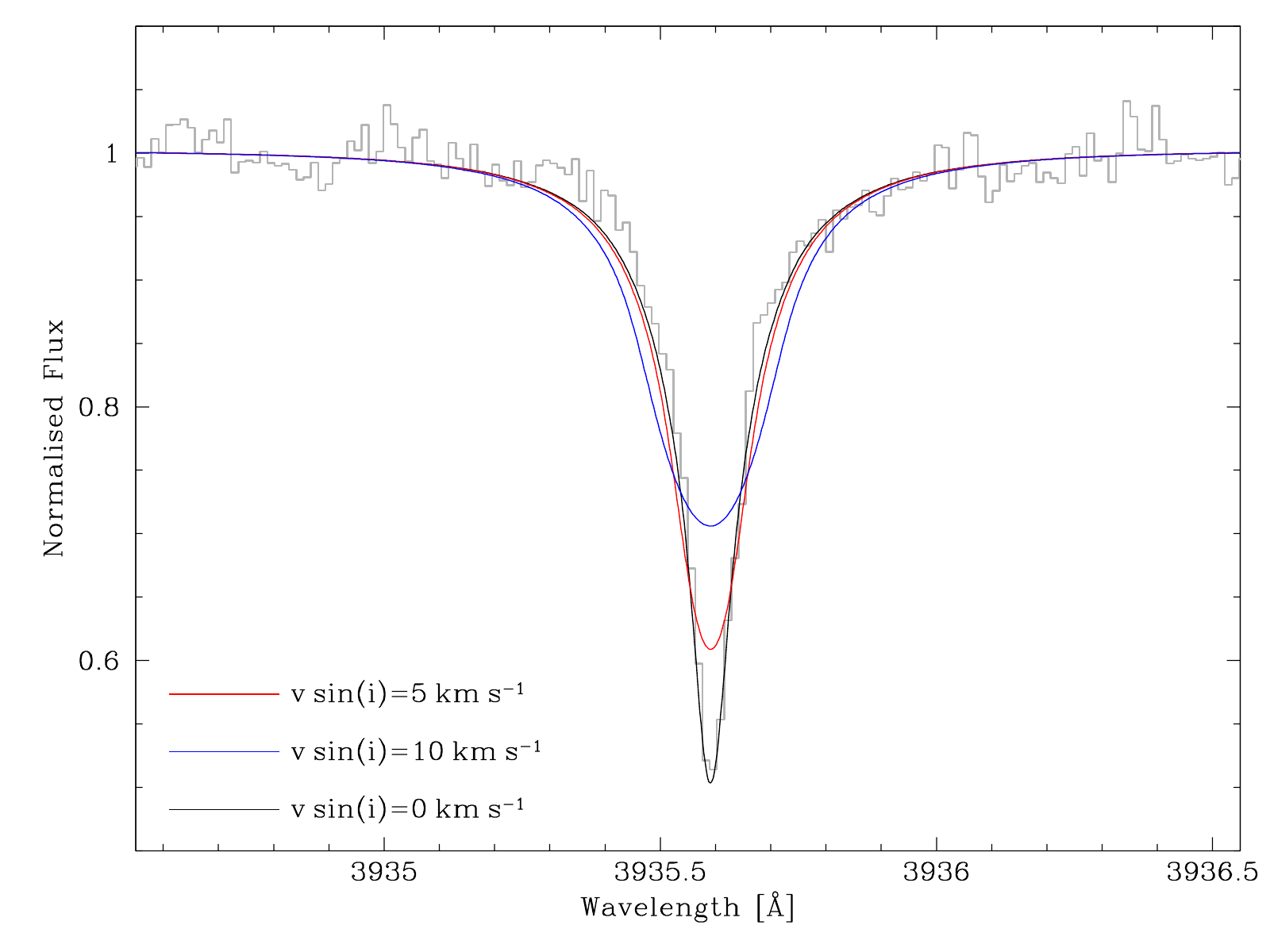}
\caption{The Ca\,K absorption in the ESPRESSO spectrum of WD\,0141$-$675 (gray) is well modelled by a non-rotating white dwarf with $\Teff=6321$\,K, $\log g=7.97$ and $\log(\mathrm{Ca/H})=-10.85$ (black line). The red (blue) lines show the model broadened by a rotational velocity of $v\sin(i)=5\,\kms$ ($v\sin(i)=10\,\kms$).}
\label{fig:vsini} 
\end{figure}

We used the extremely high resolution of the ESPRESSO spectrum to place an upper limit on the rotational velocity, $v \sin(i)\lesssim5\,\mathrm{km\,s^{-1}}$ (See Figure~\ref{fig:vsini}), corroborating earlier findings that most white dwarfs have rotation periods of many hours to several days \citep{hermesetal17-1}, too long to detect rotational broadening even in high-resolution spectroscopy \citep{koesteretal98-2, bergeretal05-2}.

\section{Probing for variability of the \ion{Ca}~K absorption line}
\label{sec:cak}

\begin{figure*}
	\includegraphics[width=1\columnwidth,trim={0.7cm 0.05cm 1.5cm 1.2cm},clip]{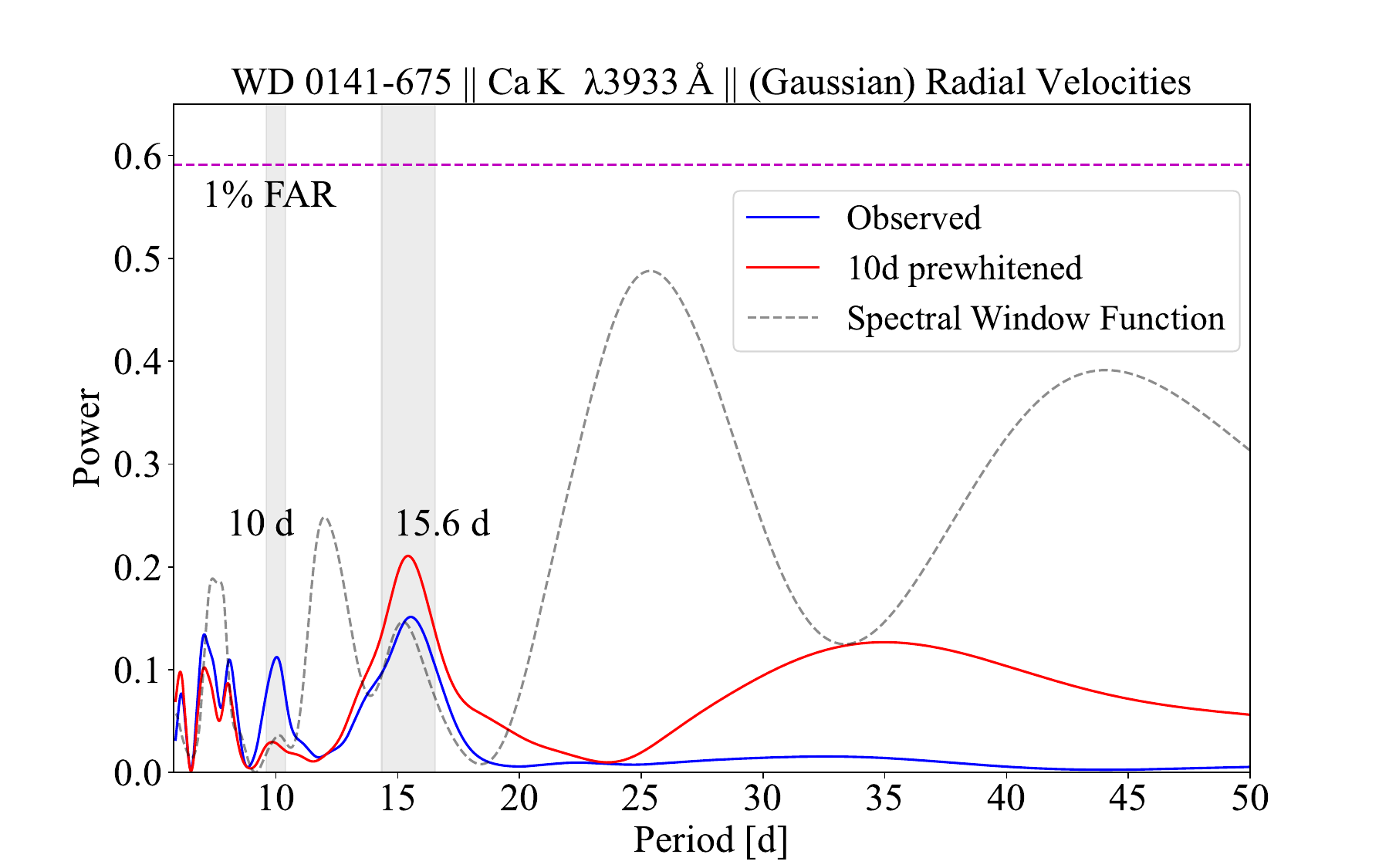}
 	\includegraphics[width=1\columnwidth]{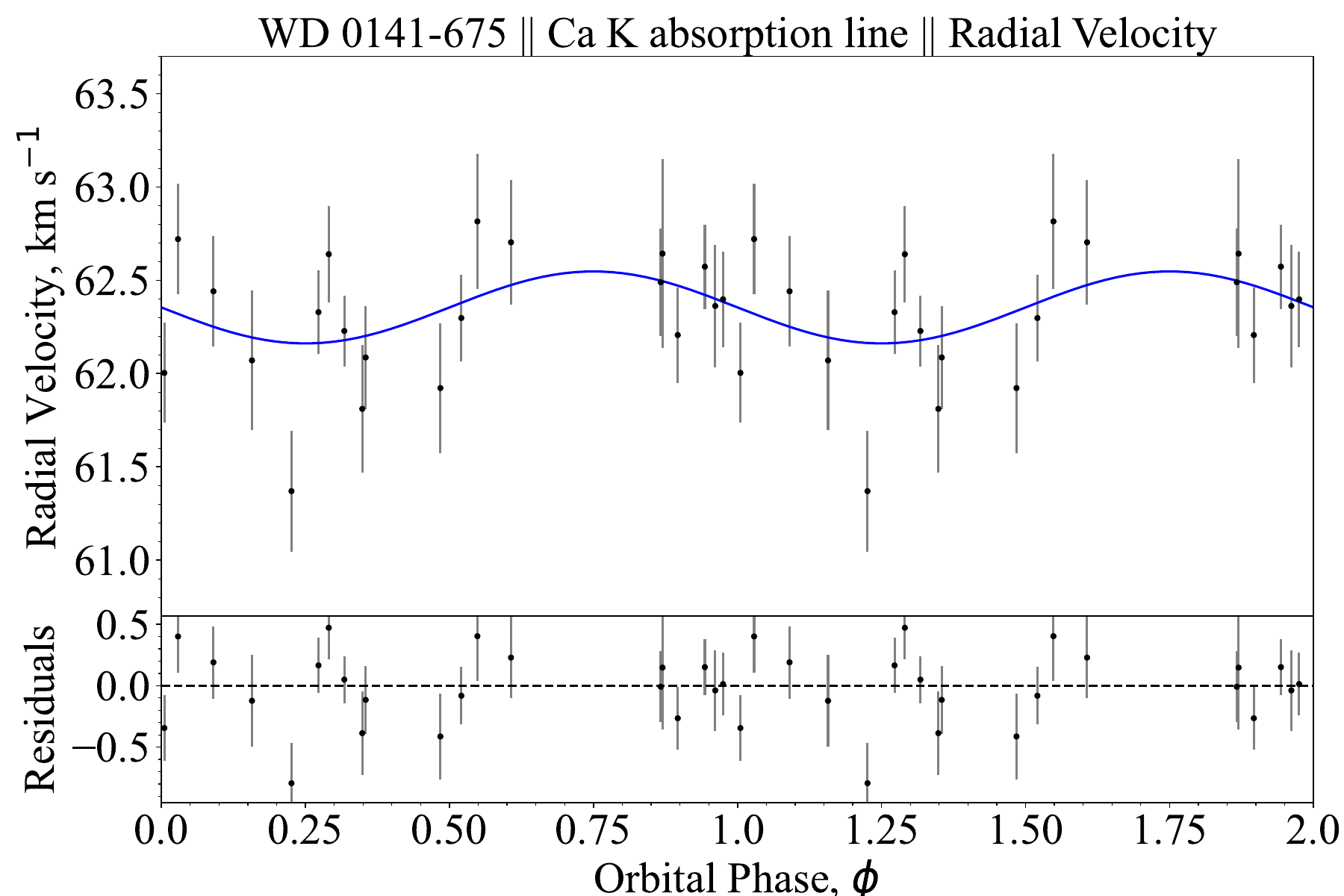}
    \caption{Left: LS periodogram of the observed radial velocities obtained for the Ca~K absorption line (in blue). The vertical gray bands highlight the strongest powers at $15.6\pm0.9$\,d and $10\pm0.4$\,d. The red curve is the LS after pre-whitening the $\simeq10$\,d signal. The magenta horizontal dashed line marks the 1~per cent False Alarm Rate. Right: Radial velocities of the Ca~K absorption line phase-folded on the $15.6\pm0.9$\,d period, with the best-fit model of a sine curve overlaid in blue; the residuals are plotted in the bottom panel.}
    \label{fig:ls_radvel_caK_observed} 
\end{figure*}

\subsection{Radial velocity analysis}
\label{sec:Radial-Velocity}

We fitted a Gaussian profile to the Ca~K absorption line to measure the radial velocity (RV) at the time of each exposure. We then computed the Lomb-Scargle periodogram from the {\sc astropy} package \citep{astropy:2022}, and calculated the False Alarm Rate (FAR) using the option based on the formulation given by \citet{baluev:2008}. \par
The RVs periodogram (left panel of Fig.\,\ref{fig:ls_radvel_caK_observed}) displays a strong peak at $P=15.6\pm0.9$\,d, roughly consistent with half the period of the astrometric planet candidate \citep[$P$=33.65,][]{gaiaetal23-1,holletal23-1}, although with a FAR of 99~per cent. Another slightly weaker signal is also detected at $10.0\pm0.4$\,d.\par 
Besides having such high FAR, each of the found signals present additional reasons for caution: The $\simeq$15\,d signal overlaps with a peak rising from the spectral window function, hinting at a possible link to the temporal sampling; and the $\simeq$ 10\,d signal coincides with a third of the lunar cycle. In Section~\ref{sec:systematics}, we explore in more detail these systematics, and argue that the $\simeq15.6$\,d signal stands as our best orbital period candidate.

Upon pre-whitening the $\simeq$10\,d period, the $\simeq$15\,d signal became stronger by $\sim$30 per cent.
And finally, pre-whitening both periods did not uncover new signals in the LS .


Figure~\ref{fig:ls_radvel_caK_observed} (right panel) shows the measured radial velocities folded with the 15.6\,d period along with a fit to a circular orbit of the form: 
\begin{equation}\label{eq:orbit}
V(t) = \mathrm{\gamma} + K \sin\left(2\pi\frac{t - t_0}{P}\right),
\end{equation}
where $\gamma$ is the systemic velocity, $K$ the amplitude, and $P$ is the orbital period. The value of $t_0$ is that of the inferior conjunction, assuming the presence of a planet. The parameters of the optimal fit are listed on Table\,\ref{tab:par-radvel}.\par 
The systemic velocity derived here, $\gamma=62.35\pm0.07\,\kms$, is somewhat lower than the average of the radial velocities reported by \citet{rogersetal24-1}, however, systematic uncertainties in the wavelength zero-point of ESPRESSO (our data), MIKE and X-Shooter \citep{rogersetal24-1} can account for this difference. The latter two instruments are not designed for stable, high-precision radial velocity studies. Indeed, the amplitude determined from the ESPRESSO spectra, $K=0.19\pm0.11\,\kms$ is well below the radial velocity precision of the MIKE data, $1\,\sigma=0.47\,\kms$ \citep{rogersetal24-1}.

We attempted the same analysis on the Ca~H line, however, due to the broad and overlapping H$\epsilon$ line, and the distortions in the flux calibration over wavelength scales comparable to the width of H$\epsilon$, we did not identify any significant variability within that line.

\begin{table}
\centering
\caption{Optimal sine fit parameters for the radial velocities.} 
\label{tab:par-radvel}
\begin{tabular}{llc}
\hline
\hline
$\gamma\, [\mathrm{km\,s^{-1}}]$        &  62.35$\pm$0.07   \\
$ \mathrm{K}\, [\mathrm{km\,s^{-1}}]$      &  0.19 $\pm$0.11\\
$t_0$\, $\left[\mathrm{MJD} \right ]$  & 60237.8 $\pm$ 1.3   \\
$\mathrm{P_{orb}}$\,$\left[\mathrm{d} \right ]$ &  15.6 $\pm$0.9 \\
\hline
\end{tabular}
\end{table}

\begin{table}
\centering
\caption{Optimal sine fit parameters for the V/R ratios.} 
\label{tab:par-vr_ratios}
\begin{tabular}{llc}
\hline
\hline
$\gamma$\textsuperscript{\textdagger}   & 0.992 $\pm$ 0.003  \\
$ \mathrm{K}$\textsuperscript{\textdagger}    &0.008 $\pm$ 0.004 \\
$t_0$\, $\left[\mathrm{MJD} \right ]$  &60236.2$\pm$2.1 \\
$\mathrm{P_{orb}}$\,$\left[\mathrm{d} \right ]$   & 16.7$\pm$ 1.0 \\
\hline

\textsuperscript{\textdagger}dimensionless& \\

\end{tabular}
\end{table}

\begin{figure*}
	\includegraphics[width=1\columnwidth,trim={0.7cm 0.05cm 1.5cm 1.2cm},clip]{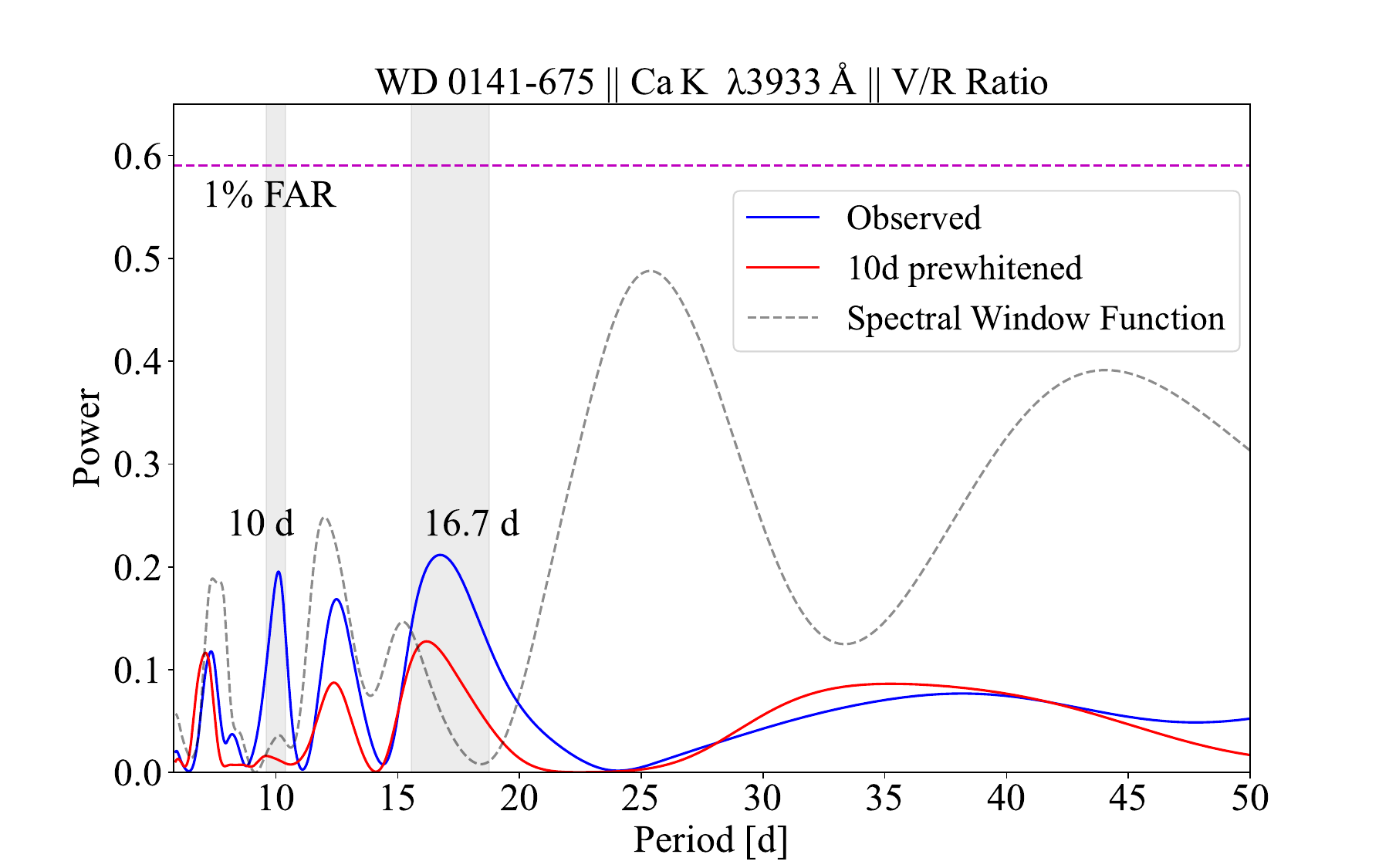}
	\includegraphics[width=1\columnwidth]{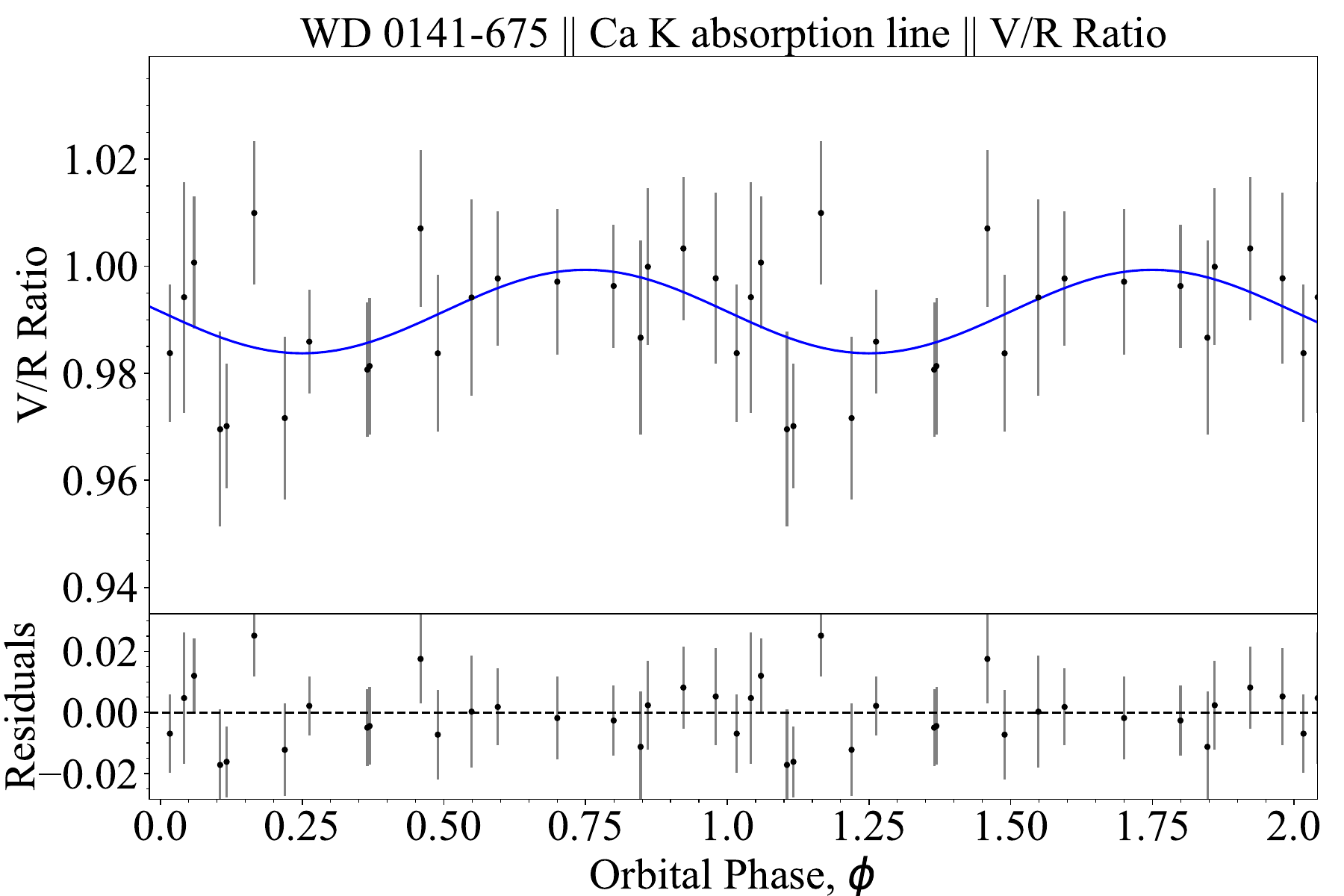}
    \caption{
    Same as Fig.\,\ref{fig:ls_radvel_caK_observed}, but for the V/R ratios of the Ca~K line measured from the ESPRESSO spectra. The strongest signals in the LS power spectrum are found at $16.7\pm0.9$\,d and $10.0\pm0.3$\,d. The right panel depicts the V/R data folded on the $16.7\pm1.0$\,d period, overlayed with the best-fit sine function. }
    \label{fig:ls_caK_observed} 
\end{figure*}

\subsection{V/R ratio Analysis}
\label{sec:vr-ratio}
We measured the violet-to-red (V/R) ratio for each ESPRESSO spectrum \citep[e.g.][]{manseretal19-1}, which is the ratio of blue-shifted to red-shifted flux, centred on the air wavelengths of the Ca~K line shifted by the systemic velocity $\gamma=62.35\,\kms$ (obtained in Section~\ref{sec:Radial-Velocity}). The LS periodogram computed from the V/R ratios (Fig.\,\ref{fig:ls_caK_observed}, left panel) contains a strong signal at $P=16.1\pm0.9$\,d, which agrees within the uncertainties with the period found in our RV analysis in Section~\ref{sec:Radial-Velocity}. We note however, that the signal has a high FAR (99~per cent). The periodogram shows two additional comparable signals at $12.5\pm0.6$\,d and $10.0\pm0.3$\,d. \par
In contrast to the RV analysis in Section~\ref{sec:Radial-Velocity}, pre-whitening the $\simeq$10\,d signal decreased the power of both the $\simeq16$\,d and the $\simeq12$\,d signals. Further pre-whitening of the $\simeq16$\,d period did not reveal new signals in the LS.\par
The results of the V/R analysis are sensitive to the interval adopted in the integration of the V/R ratio, whereby wider intervals favoured the strength of the shorter periods, while narrow intervals increased the strength of the  $\simeq16\,$d period. Our final choice was an interval of $-35$ to $35\,\kms$ around the core of the line, which yields all three signals with a comparable strength. \par
We phase-folded the data with the $16.7\,$d-period and fitted a sine function to the data. The best fit is shown in the right panel of Fig.\,\ref{fig:ls_caK_observed}, and the corresponding parameters are listed on Table\,\ref{tab:par-vr_ratios}. As in the RV measurement discussed above, the periodic signal detected in the V/R data is consistent with half of that reported for the retracted planet candidate ($P=33.65$\,d). On the other hand, it is worth noting that the amplitude modulation $K$ and systemic offset $\gamma$ are dimensionless parameters in this analysis, and are therefore not comparable to the parameters obtained in the RV study from Section~\ref{sec:Radial-Velocity}.\par
\begin{figure}
	\includegraphics[width=1\columnwidth,trim={0 3.1cm 0 4cm},clip]{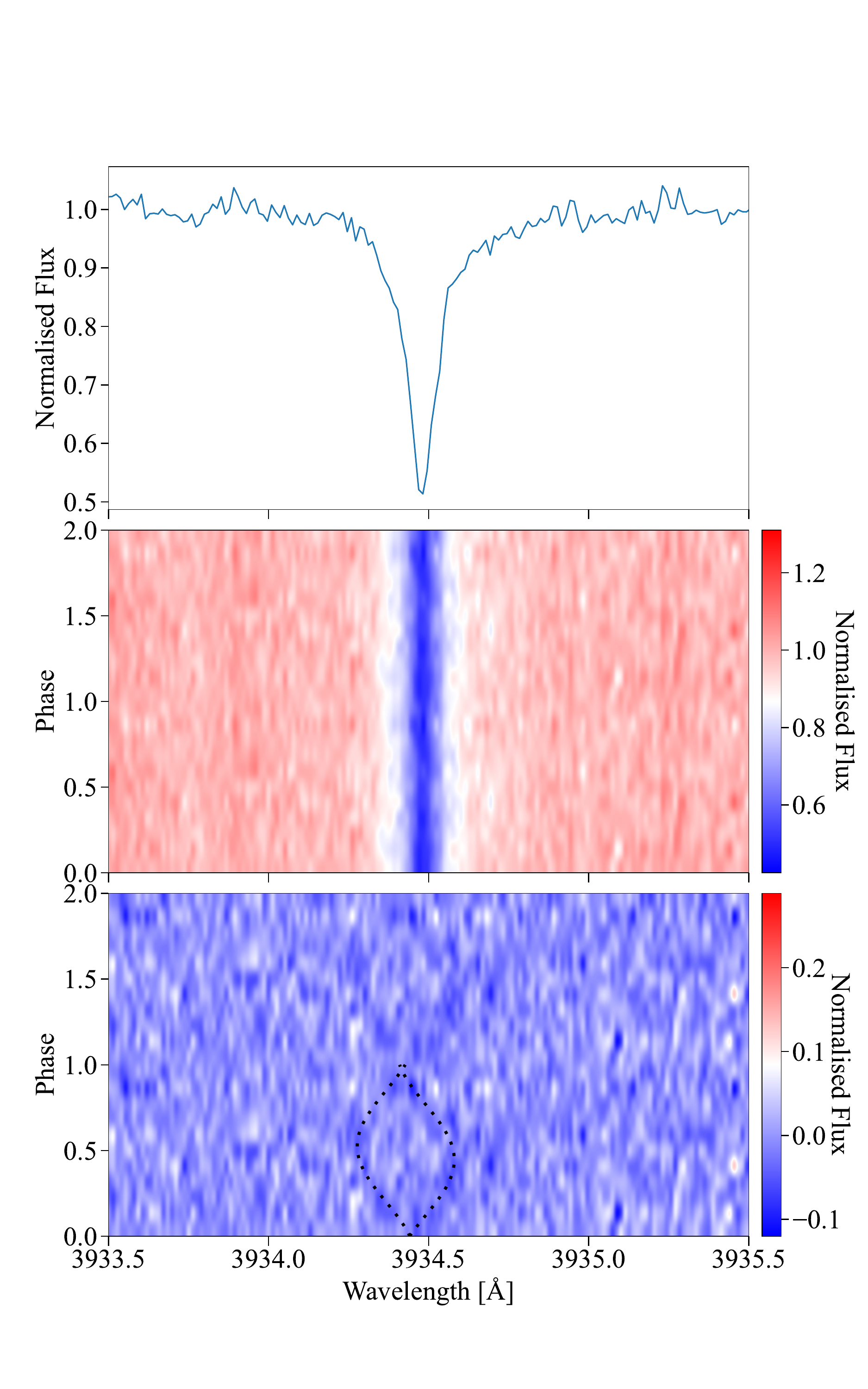}
    \caption{Top panel: Continuum normalised average spectrum. Middle panel: Trailed spectrogram constructed from the the time-resolved ESPRESSO data after normalising by the continuum. The individual spectra were folded on the $16.7\pm1.0$\,d period and binned in 12 evenly spaced phase bins. We show two phase cycles for visual purposes. Bottom panel: Trailed residuals after subtracting the average spectrum from each exposure. Two sine waves emerge in the residuals which are over-plotted by black dotted sine curves over the first orbit. The latter do not represent a fit to the data but are included only as a guide to the eye.  }
    \label{fig:trail_caK_observed} 
\end{figure}

To visualise the variability of the spectra along the orbit, we first computed the average spectrum of all the ESPRESSO observations, subtracted that average from each individual spectrum, and then phase-folded the data on the $\simeq16$\,d period, with the result displayed as a phase-trailed spectrogram in Fig.\,\ref{fig:trail_caK_observed}. Two tentative sine wave patterns are present in the residual spectrogram (bottom panel of Fig.\,\ref{fig:trail_caK_observed}), but a spurious origin cannot be discarded without more observations. \par
We attempted the same V/R analysis on the \ion{Ca}~H line, but similar to Section~\ref{sec:Radial-Velocity}, the overlapping broad H$\epsilon$ absorption impeded a clean V/R measurement.

\begin{figure}
	\includegraphics[width=1\columnwidth,]{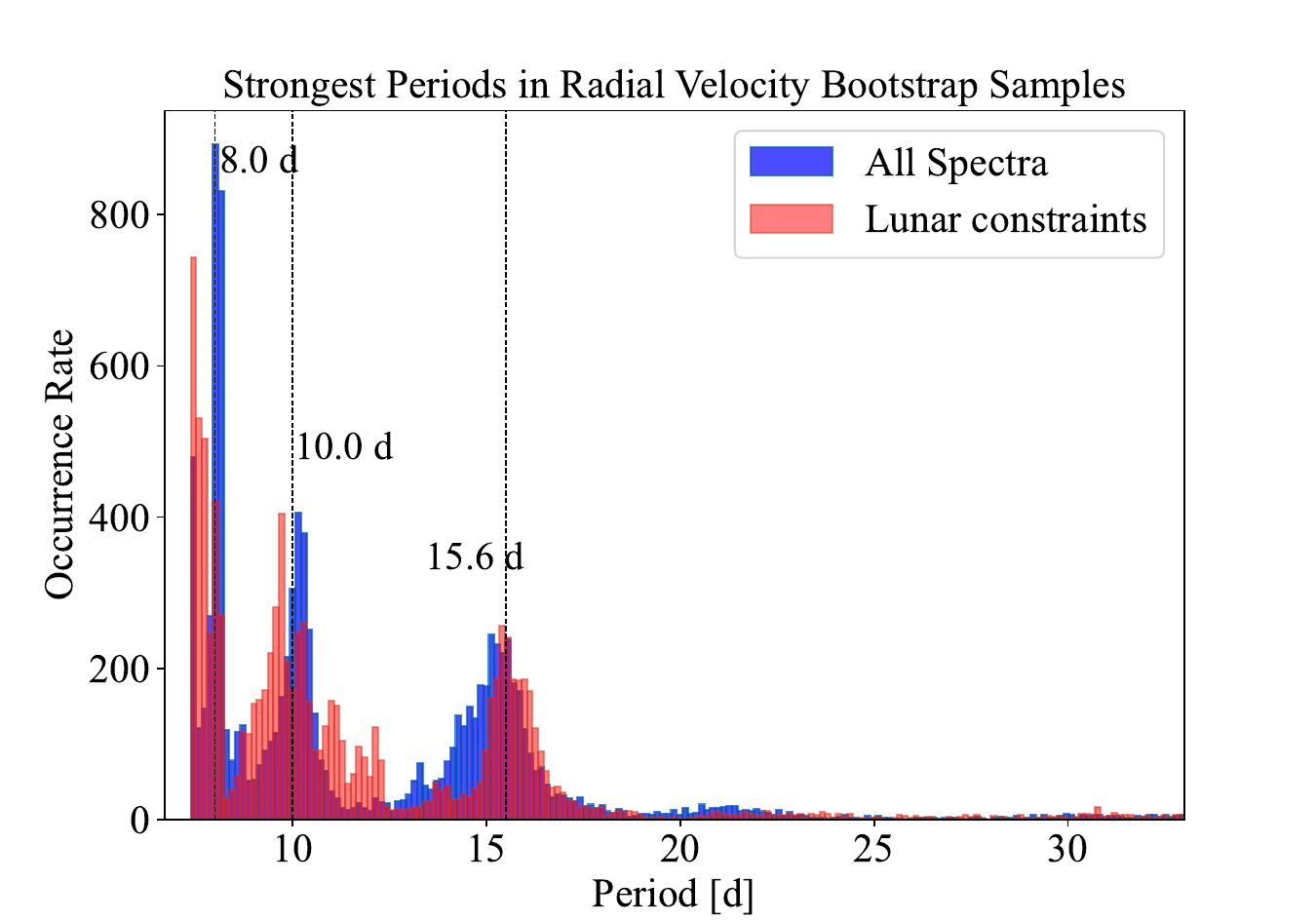}
	\includegraphics[width=1\columnwidth]{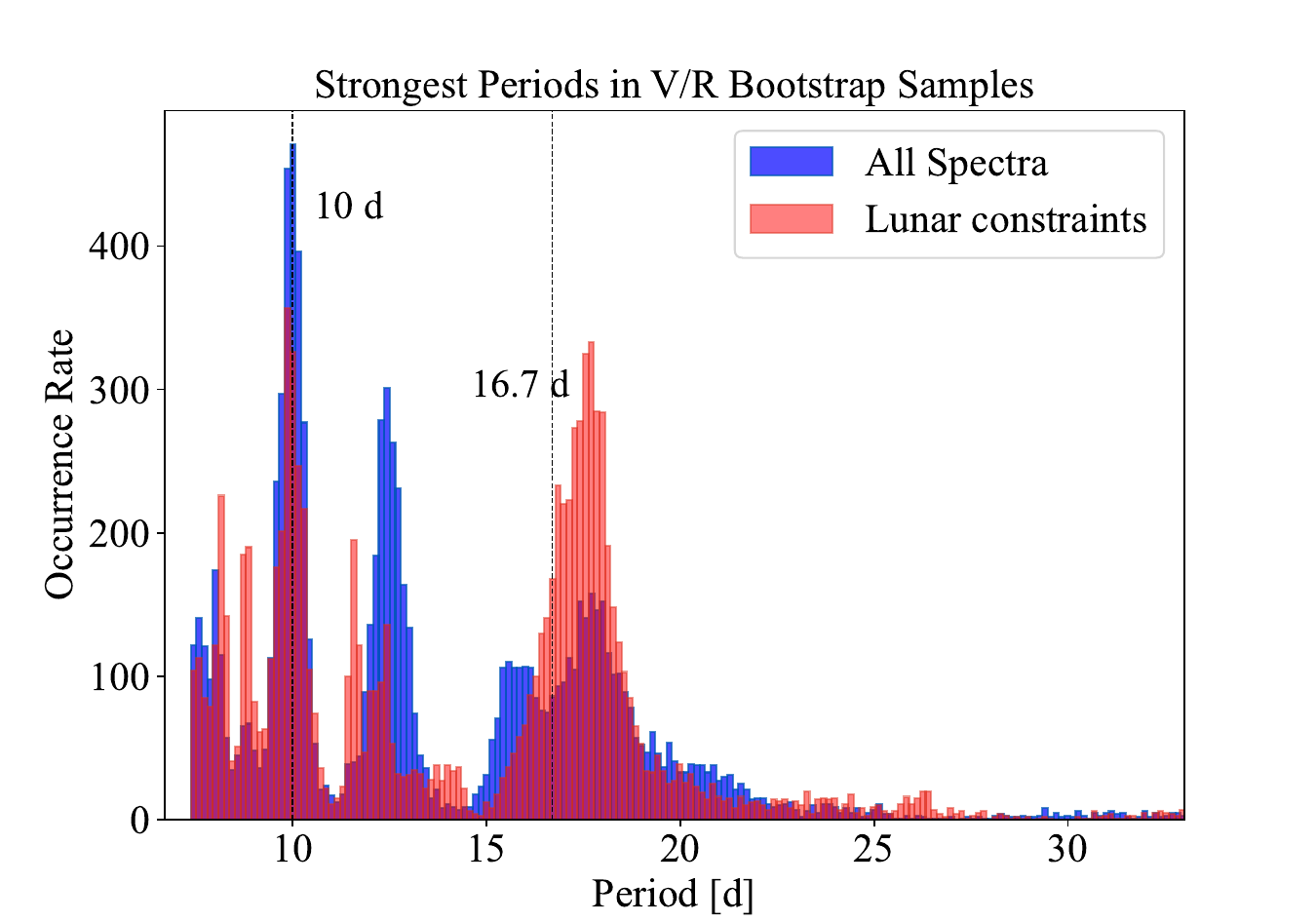}
    \caption{
    Top: Histogram of the strongest periods found in the 10\,000 radial velocity bootstrap samples. The blue histogram shows the results of the complete set of spectra. For the red histogram we only considered spectra satisfying the imposed constraints on moon angular separation ($>65^{\circ}$), fraction lunar illumination ($<0.85$), and S/N ($>10$) around \ion{Ca}~K. The vertical dashed lines indicate the position of the strongest periods found in the observed data. Bottom: Same as top panel but for the V/R ratio measurements.}
    \label{fig:histograms} 
\end{figure}

\begin{figure}
	\includegraphics[width=1\columnwidth]{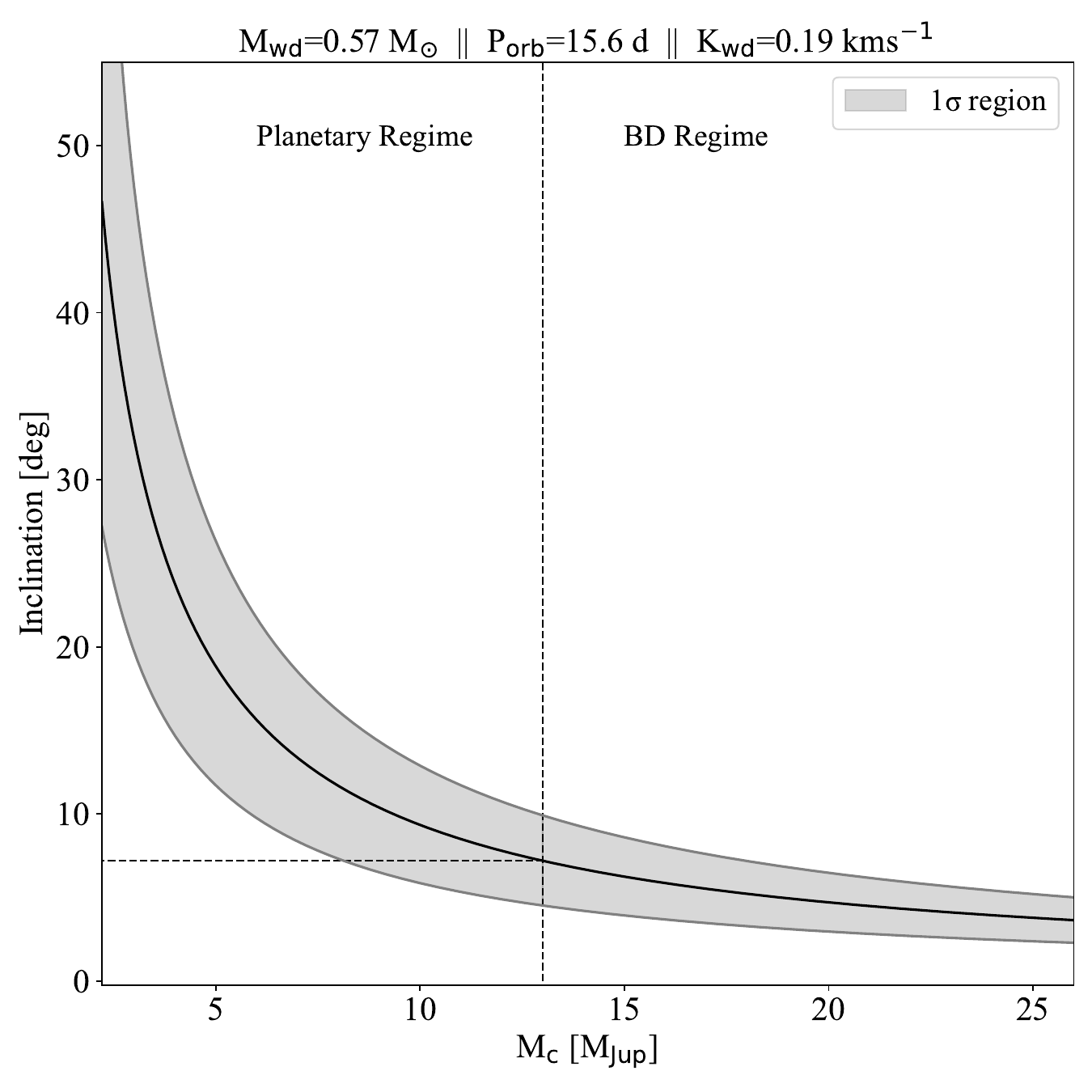}
    \caption{Mass function of a putative companion to WD\,0141--675, calculated from the orbital parameters measured from the radial velocity analysis (Section~\ref{sec:Radial-Velocity}, Table\,\ref{tab:par-radvel}). The black line represents the mass function of the central values corresponding to $P_\mathrm{orb}=15.6\pm0.9$\,d and $K_\mathrm{wd}=0.19\pm0.11\,\kms$. The grey region shows the  constraints on the inclination and mass of a possible companion within the $1\sigma$ region implied by the uncertainties in our measurements. The dashed vertical line indicates the $13\,\mathrm{M_{Jup}}$ threshold above which the companion could also be a brown dwarf. 
    Assuming a planetary companion, the mass boundary implies a lower limit on the inclination of $i=7\pm4^\circ$, indicated by the dashed horizontal line.
}
\label{fig:mass-inclination} 
\end{figure}

\section{Discussion}
\subsection{Systematics Test}
\label{sec:systematics}
We explored two possible systematic origins for the periodic signals arising in the radial velocity and V/R measurements: The temporal sampling of the ESPRESSO spectra and the lunar cycle. For this purpose we produced 10\,000 bootstrap samples of our measurements for which we computed LS periodograms, recording the period with the strongest power in each sample. We then repeated the experiment imposing constraints on the angular separation from the Moon ($>65^{\circ}$), the fraction lunar illumination (FLI$<0.85)$, as well as on the S/N ($>10$) around \ion{Ca}~K. We show the histograms of the most frequent strongest periods found in the bootstrap samples in Fig.\,~\ref{fig:histograms}. When using the complete set of spectra the same signals dominating the LS periodograms in Sections~\ref{sec:Radial-Velocity} and \ref{sec:vr-ratio} are those appearing most frequently in the bootstrap histograms. The only exception is the $\simeq8.0$\,d signal in the radial velocity histogram, which is consistent to the harmonic in frequency of the $\simeq$16\,d signal. These results suggest that the temporal-sampling is not the cause of the detected signals.

When imposing the lunar and S/N constraints on the radial velocities (top panel of Fig.~\ref{fig:histograms}), the overall structure remains mostly unaffected. However,
in the case of the V/R measurements the signal at $16.\pm1.0$\,d increases its occurrence rate at the expense of a decrease in the occurrence of the $\simeq $10 and 12\,d signals.

This marginally hints at sky background contamination as a possible origin for these shorter signals. For this reason, we argue that the signal at $\simeq16$\,d is the best orbital period candidate; but we still issue caution, as further observations are needed to sustain such assessment.\par

\subsection{Implications on the Mass Function}
In Section\,\ref{sec:Radial-Velocity}, we tentatively detected a radial velocity variation in the Ca~K line with $P=15.6\pm$0.9\,d and an amplitude of  $K=0.19\pm0.11\,\kms$. Furthermore, a consistent period within the errors of $P=16.7\pm1.0$\,d was also found in the V/R analysis (Section~\ref{sec:vr-ratio}). This period agrees within uncertainties with that of the astrometric period candidate \citep[$P=33.65$\,d,][]{gaiaetal23-1,holletal23-1}\par
Taking the RV parameters at face value and interpreting them as the circular orbital motion of the white dwarf resulting from a close-in planet leads to the following mass function for the putative companion:
\begin{equation}
f(M_\mathrm{c})=\frac{PK_\mathrm{wd}^3 }{2\pi G}=\frac{({M_\mathrm{c}}\sin{(i)})^3}{(M_\mathrm{wd}+M_\mathrm{c})^2}=1.20\times10^{-5}\,\mathrm{M_{Jup}}
\label{eq:mfunc-inc}
\end{equation}

\noindent
where $G$ is the gravitational constant and $M_\mathrm{wd}=0.57\pm0.02$\,$\mathrm{M_{\odot}}$ the mass of the white dwarf as reported by \citet{rogersetal24-1}. This mass function constrains the possible values for the inclination ($i$) and the mass of the companion ($M_\mathrm{c}$), as illustrated in Fig.\,\ref{fig:mass-inclination}. For an orbital inclination $i\lesssim7^\circ$ the companion would have a mass in the brown dwarf regime, whereas $i\gtrsim7^\circ$ allows only for a planetary mass companion. The companion mass reported by \textit{Gaia}, $\mathrm{M_{c}}=9.26^{+2.64}_{-1.15}\,\mathrm{M_{Jup}}$, yields an orbital inclination of $i=10.20^{\circ}$ in our analysis, which is in stark disagreement with the inclination of $i=87.0\pm4.1^\circ$ inferred from Monte Carlo resampling by \citet{gaiaetal23-2}. However, we remind the reader that the astrometric solution for WD\,0141--675 has been retracted by the \textit{Gaia} team. We also note that while the available \textit{Spitzer} and \textit{WISE} photometry rules out a brown dwarf companion (see section\,3.3 of \citealt{rogersetal24-1} for a detailed discussion), a more stringent photometric limit on the existence of a planetary companion could be derived from deep \textit{JWST} infrared observations\footnote{The approved \textit{JWST} program GO~3652 intended to obtain these observations, but was withdrawn in the wake of the retraction of WD\,0141--675. We argue that a better constraint on the nature of a possible companion to WD\,0141--675 remains of interest.}.

In summary, interpreting the radial velocity variation tentatively detected in the ESPRESSO data as the reflex motion of the white dwarf, in conjunction with the existing infrared data on WD\,0141--675, implies possible solutions for a planetary companion with a lower limit of $\sim7^{\circ}$ on the orbital inclination. We stress, however, that the radial velocity signal in the ESPRESSO data is not sufficiently significant to claim any detection of a companion.  

\section{Conclusions}
\obj\ was reported as hosting the first astrometrically detected white dwarf planet candidate \citep{gaiaetal23-1,holletal23-1}, but was later retracted as a false positive. We have obtained time-resolved high-resolution spectroscopy of \obj\ with ESPRESSO, an instrument which allows highly accurate velocity shift measurements \citep[e.g.][]{pasquinietal-2023}. Adopting the atmospheric parameters from the photometric solution of \citet{rogersetal24-1}, we obtained a Ca abundance of $\log(\mathrm{Ca/H})=-10.75\pm0.05$\,dex, in good agreement with the values reported by these authors. The narrow width of the Ca~K lines yielded an upper limit on the rotational broadening of $v\sin{i}\simeq5\,\kms$. This is suggestive of a slow rotator, which is a common result in spectroscopic determinations of rotating rates in white dwarfs \citep{bergeretal05-2}. 

Analysing the radial velocities and the V/R ratio of the Ca~K line both result in a tentative detection of variability with a period of $\simeq15-16$\,d, which cannot be attributed to the temporal sampling. Phase-folding the ESPRESSO spectroscopy on such signal exhibits weak periodical variability in the morphology of
\ion{Ca}~K near its core. Fitting the radial velocities with a circular orbit results in an amplitude of $K=0.19\pm0.11\,\kms$.   
 Interpreting this variation as the (circular) reflex motion of the white dwarf implies that possible solutions to the mass function exist within the planetary mass regime. Therefore, when discarding  a brown dwarf companion from existing infrared photometry \citep{rogersetal24-1} a lower limit on the orbital inclination of $i\gtrsim7^\circ$ is established. \par

Whereas we stress that the measured radial velocity variation is not statistically significant, our analysis demonstrates that ESPRESSO observations are entirely sensitive to detecting planets on close-in orbits around white dwarfs. We encourage additional ESPRESSO observations, as well as deep infrared photometry to increase the dynamic and photometric constraints on a putative companion to WD\,0141--675.

\section*{Acknowledgements}

We thank James Munday for useful discussions on radial velocity methods. We would also like to thank the anonymous referee, whose comments helped improve the content of this article. This project has received funding from the European Research Council (ERC) under the European Union’s Horizon 2020 Framework Programme (grant agreement no. 101020057). This research was in part funded by the UKRI (Grant EP/X027562/1)
Research based on observations collected at the European Organisation for Astronomical Research in the Southern Hemisphere under ESO programme(s) 112.2604.001.

\section*{Data Availability}
The ESPRESSO data are available from the ESO archive.



\bibliographystyle{mnras}
\bibliography{mnras} 



\appendix

\section{Log of the observations}

\begin{table}
\centering
\caption{\label{t-obslog}Log of the ESPRESSO spectroscopy, with the date referring to the start of the exposure. Each spectrum was taken with a 3144\,s exposure time. The lunar distance refers to the angular separation from the moon in degrees.}
\begin{tabular}{cccccc}
\hline
Date & Time & MJD & Lunar  & FLI & S/N \\
& &  &  Dist.  &  & \\
\hline
2023-10-20 & 03:13:49 & 60237.134604 & 74 & 0.3 & 20.9 \\
2023-10-21 & 02:16:07 & 60238.094533 & 69 & 0.4 & 21.5 \\
2023-10-25 & 06:03:14 & 60242.252246 & 64 & 0.8 & 20.1 \\
2023-10-29 & 02:32:04 & 60246.105605 & 84 & 1.0 & 20.3 \\
2023-11-05 & 03:57:32 & 60253.164957 & 118 & 0.5 & 20.2 \\
2023-11-11 & 01:32:23 & 60259.064163 & 105 & 0.1 & 19.9 \\
2023-11-13 & 01:54:27 & 60261.079487 & 94 & 0.0 & 17.5 \\
2023-11-14 & 01:52:00 & 60262.077783 & 88 & 0.0 & 14.2 \\
2023-11-19 & 01:41:56 & 60267.070795 & 63 & 0.3 & 14.1 \\
2023-11-26 & 00:39:32 & 60274.027459 & 87 & 1.0 & 25.8 \\
2023-12-05 & 00:30:18 & 60283.021052 & 117 & 0.5 & 22.2 \\
2023-12-06 & 00:46:26 & 60284.032248 & 115 & 0.4 & 18.1 \\
2023-12-07 & 01:55:34 & 60285.080263 & 111 & 0.3 & 19.5 \\
2023-12-08 & 00:52:07 & 60286.036197 & 107 & 0.2 & 16.3 \\
2023-12-09 & 01:47:59 & 60287.074998 & 102 & 0.2 & 12.0 \\
2023-12-10 & 03:21:44 & 60288.140102 & 96 & 0.1 & 13.9 \\
2023-12-11 & 03:30:44 & 60289.146346 & 89 & 0.0 & 19.3 \\
2023-12-12 & 01:11:57 & 60290.049968 & 84 & 0.0 & 16.8 \\
2023-12-16 & 01:28:59 & 60294.061797 & 64 & 0.1 & 18.0 \\
2023-12-20 & 02:23:25 & 60298.099596 & 69 & 0.5 & 19.0 \\
\hline
\end{tabular}
\label{tab:your_label}
\end{table}

\bsp	
\label{lastpage}
\end{document}